# How to Motivate and Engage Generation Clash of Clans at Work? Emergent Properties of Business Gamification Elements in the Digital Economy

Nicholas Dacre [a,*], Panos Constantinides [a], & Joe Nandhakumar [a]

[a] Warwick Business School, University of Warwick, Coventry, CV4 7AL, UK.
[*] Corresponding Author: nicholas.dacre@mail.wbs.ac.uk

## Abstract

Organisations are currently lacking in developing and implementing business systems in meaningful ways to motivate and engage their staff. This is particularly salient as the average employee spends eleven cumulative years of their life at work, however less than one third of the workforce are actually engaged in their duties throughout their career. Such low levels of engagement are particularly prominent with younger employees, referred to as Generation Y (GenY), who are the least engaged of all groups at work. However, they will dedicate around five cumulative years of their life immersed playing video games such as 'Clash of Clans', whether for social, competitive, extrinsic, or intrinsic motivational factors.

Using behavioural concepts derived from video games, and applying game design elements in business systems to motivate employees in the digital economy, is a concept which has come to be recognised as Business Gamification. Thus, the purpose of this research paper is to further our understanding of game design elements for business, and investigate their properties from design to implementation in gamified systems. Following a two-year ethnographic style study with both a system development, and a communication agency largely staffed with GenY employees, findings suggest properties in game design elements are emergent and temporal in their instantiations.

**Keywords**: Business Gamification, Motivation, Digital Economy, Leaderboard, Clash of Clans, Enterprise Gamification, Gamification, Engagement, Self-Continuity, Employee, Gen Y, Workforce, Staff, Millennials.

# Introduction

Companies striving to embrace the digital economy (Skilton, 2015) have historically committed considerable resources into developing their business systems towards greater efficiency (Ciborra & Hanseth, 1998; Strader, Lin, & Shaw, 1998). This is particularly salient as organisations transition from relying on incumbent platforms, such as legacy ERP or CRM infrastructures as a source of competitive advantage towards implementing new and innovative internal business applications (Herzig, Ameling, & Schill, 2012). In this context, built on an architecture of dispersed, highly interconnected, always-on systems such as cloud-based corporate platforms, the digital economy affords businesses innovative opportunities to disrupt organisational processes and enhance their competitive advantage (Briscoe & Marinos, 2009; Skilton, 2015). However, whilst focusing on elements of operational functionality (Ciborra, 2000), organisations have largely neglected human-elements of motivation and engagement in the development and implementation of their business systems.

Considering that two thirds of employees are either disengaged or actively disengaged in their work activities (O'Boyle & Harter, 2014), companies must also consider the challenges of staff motivation and engagement across their business systems. One novel approach to applying these dimensions in the digital economy, is by leveraging certain highly engaging game design elements and game mechanics typically derived from video games (Thom, Millen, & DiMicco, 2012). In order to further understand the motivational properties of these game design elements, and what this may signify for business systems, in the following section we focus on one of the most successful games currently in circulation called 'Clash of Clans' and the significance with its users.

## Generation Clash of Clans

Clash of Clans originally developed by Finnish company Supercell (Carayannis & Rakhmatullin, 2014), is both outperforming and outranking in terms of downloads and active users, most other titles in the gaming industry, including well renowned Candy Crush Saga (Erturkoglu, Zhang, & Mao, 2015). The game's success is such that it has displaced traditional Massively Multiplayer Online Game (MMOG) behemoth World of Warcraft which held a record breaking 12 million users at its peak (Lee et al., 2011). In contrast, Supercell have an estimated 29 million active daily users across their gaming platforms, which in turn generate around $5 million in revenue each day (Cheng, 2014). This accessible and yet highly addictive strategy game represents a confounding success of mobile and gaming innovation in the digital economy.

The largest segment of players in Clash of Clans comprise of Generation Y (GenY) participants (as shown in Figure 1). These individuals are highly adept with nascent technology, have grown to expect instant feedback, and espouse greater levels of self-determination (Connor et al., 2008; MacLeod & Clarke, 2009). Players immersed in Clash of Clans will also elect to dedicate anywhere between 30 minutes and 6 hours of gaming per day, whilst trying to earn virtual trophies, climb the leaderboard, and engage in battle quests (Supercell, 2014).

However, these figures are in stark contrast to the levels of engagement and motivation of analogous employees in the workplace (O'Boyle & Harter, 2014). Thus, the game design elements employed to captivate and engage users in Clash of Clans, should be of significance for organisations wishing to engage their employees across their business systems in the digital economy.

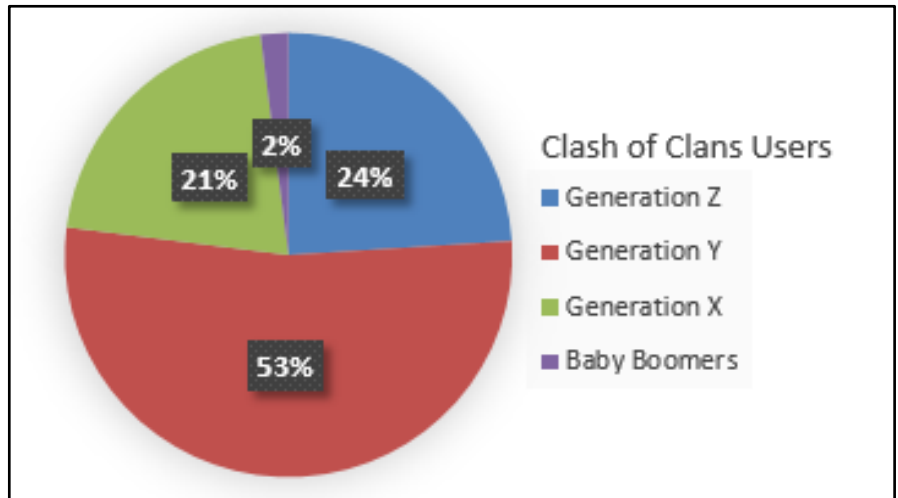

Figure 1: Clash of Clans by Gen. Identifier (adapted from newzoo data explorer)

**The Disengaged Workforce**

An overwhelming two thirds of staff are either not engaged or actively disengaged whilst at work costing the UK economy between £52 and £70 billion in lost productivity each year (O'Boyle & Harter, 2014). Such is the problem that the UK Secretary of State for Business, initiated an investigation which lead to the 2009 MacLeod 'Engagement for Success' report (MacLeod & Clarke, 2009). One particular aspect of the document outlined that younger employees "want more out of work than simply a wage packet at the end of the week" (ibid, p.29). Furthermore, GenY workers represent the least engaged of all the groups in the workplace (Adkins, 2015). This is particularly problematic for organisations that rely on a growing number of younger graduate employees who are less "willing to abandon that desire for self-determination when they enter work" (MacLeod & Clarke, 2009, p. 29).

As illustrated in Clash of Clans, GenY will elect to dedicate a considerable amount of time actively participating across a number of interactive, social, collaborative and competitive activities whilst immersed in video games (McGonigal, 2011; Supercell, 2014). This poses a distinct challenge for companies on how to design and implement business systems that may motivate and engage GenY staff. In response to these challenges there is increasing evidence that organisations such as Delloitte, Capgemini, IBM, and SAP for example, are introducing game design elements into their business systems (El-Masri et al., 2015; Hamari, Koivisto, & Sarsa, 2014; Herzig et al., 2012). Ultimately, towards motivating and engaging their employees, particularly GenY, across a multitude of behavioural dimensions. This nascent process has come to be recognised as the 'Gamification' of internal employee business systems towards greater motivation and engagement across the enterprise.

In this section we have identified game design elements as particularly salient for business systems in order to motivate and engage, an otherwise disengaged generation of employees. Now a discussion on the conceptual background and understanding of gamification is presented. Particular relevance is focused on the concepts of business gamification and self-determination theory, followed by literary evidence of application in practice. The research and approach section outline the applied methodology followed by the analysis and findings. Finally, the contributions are presented.

## Conceptual Background

Gamification refers to the idea of incorporating a variety of game design elements, mechanics, dynamics, and behavioural approaches typically derived from video games to non-game contexts (Burke, 2014; Deterding et al., 2011; Hamari, Huotari, & Tolvanen, 2015; Kapp, 2012). Implementing games outside of traditionally recognised leisure activities, such as in a work environment, in itself is not a new phenomenon. One of the most notable applications of this approach to date has been through 'Serious Games'. However, the concept of gamification should not be confused with serious games. Although some similarities can be drawn between serious games and gamification, and the terms in some scholarly articles have been used interchangeably (Richter, Raban, & Rafaeli, 2015), serious games and gamification however are not one and the same (Landers et al., 2015).

Serious games present an embodiment of a game, as in the virtual representation often delivered as a 3D construct built on the same architecture as video games (such as the Unity game engine) which encompass levels of interactivity through a graphic user interface (Petridis et al., 2010). Where a video game might be centred on leisurely pursuits (Wood, Griffiths, & Parke, 2007), serious games however are generally developed for educational and training purposes (Michael & Chen, 2005). Gamification, in contrast to serious games, is not generally supported by a fully stimulatory environment developed through a game engine. Instead it proposes the adoption of game design elements typically derived from video games and applied to non-game contexts (Deterding et al., 2011), towards eliciting customer or employee engagement (Herger, 2014; Zichermann & Cunningham, 2011).

### Business Gamification

Whilst gamification has been applied in a variety of organisational settings (Deterding, 2012; Khatib et al., 2011; King et al., 2013; Monu & Ralph, 2013; Singh, 2012), the concept itself can be delineated as either customer-focused or business-focused.  Where a customer-focused approach to gamification is adopted, game design elements are normally used to engage consumers across a number of marketing approaches (Huotari & Hamari, 2012; Paharia, 2013; Zichermann & Cunningham, 2011). However, where gamification is used in a business environment, this is referred to as 'Business Gamification' or 'Enterprise Gamification' and is typically focused on engaging and motivating employees (Herger, 2014; Mollick & Rothbard, 2013; Penenberg, 2013; Reeves & Read, 2013). In this context, gamification enables organisations focused on business issues, on providing meaningful methods in which to apply a range of motivational and engaging game design elements through their enterprise systems (Herzig et al., 2012; Rauch, 2013; Thom et al., 2012).

The result and success of this type of gamification can vary depending on how these elements are applied across business systems, and the context in which they are adopted (Thom et al., 2012). One such factor of success is by establishing greater levels of self-determination through the application of intrinsic and extrinsic motivational factors (Hamari et al., 2014; Muntean, 2011; Ryan & Deci, 2000).

**Self-Determination Theory and Gamification**

As outlined in the MacLeod report (2009), self-determination is a particularly important dimension for GenY in the workplace. The concept of self-determination has a long research history, predominantly through the work of Ryan & Deci (2000) attributing dimensions of Competence, Relatedness and Autonomy as basic universal human requirements towards greater self-continuity (Sani, 2010). Essentially, these dimensions seek to provide levels of mastery, connection and independence in supporting basic human psychological needs, whilst not necessarily disconnecting individuals from collaborative activities (Deci & Vansteenkiste, 2004).

Self-determination theory has more recently been the focus of human engagement and motivation research through the exploration of intrinsic and extrinsic values in gamified systems (Hamari et al., 2014; Nicholson, 2012). Engagement and motivation in self-determination theory state that, greater the levels of intrinsic motivation, as in an internalised or emotional stimulus, the greater the pull towards completing an activity (Ryan & Deci, 2000). In contrast, where activities are inherently driven by extrinsic motivation, these may motivate in the short-term but can also have a detrimental effect on the long term intrinsic motivational factors of an activity. Having reviewed the conceptual background to business gamification, we now outline our methodological approach and the context of our empirical research.

# Research Approach

This paper draws on ethnographic data collected over two years from both a system development, and a communication agency with a largely GenY workforce. From the onset of the research process, the agency was operating in a market with progressively diminishing margins, and required new and innovative ways to motivate and engage its employees towards increasing its competitive advantage. The lead researcher was able to observe and capture both the development and implementation of this gamified business system throughout the two years of fieldwork, during which the following key research question was examined; If gamification is the use of game design elements in a non-game context, what are the properties of game design elements in gamified business systems?

The process of the project endeavoured both the system developer and the communication agency to work together in developing various strategies, game mechanics and game design elements. These would be implemented into a gamified business system that would not only facilitate the effective support of key business processes for the agency, but also provide a highly innovative and engaging platform for their employees.

## Analysis and Initial Findings

In this analysis we focus on the emergent properties of one of the most prominent game design elements in the agency's gamified business systems. We use the concept of self-continuity (Sani, 2010) to specifically focus on a distinct game design element which embodies the reflection of the 'self' over time. This is particularly important as self-continuity elements can be recognised by individual achievements, badges, trophies, or evolutionary aspects of an online profile in a gamified business system or video game. For example, one of the most prominent game design elements of self-continuity in the Clash of Clan video game, is through the leaderboard system (as show in Figure 2).

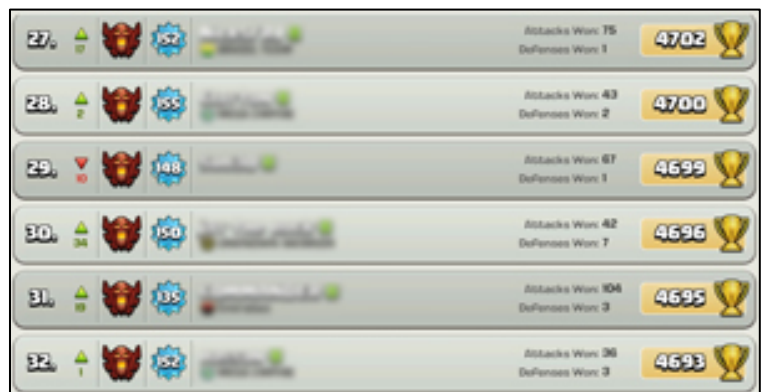

Figure 2: Clash of Clans Leaderboard System

Self-continuity in this context, represents the projection of the person's various achievements, goals and online profile over time. Social psychologist, Professor Sani states that "we have a sense of self-continuity because inside us there truly is something that corresponds to a continuous self, something that is the subject of all our experiences" (2010, p. 1). In the context of our empirical study we use this interpretation of self-continuity to identify the agency's Leaderboard system as a particularly salient game design element, and its significance for individual employees perpetuated with the self over time.

### Leaderboard

The agency's Leaderboard system offers a visual representation of the top tier of system users identified as highflyers. Every time an employee logs into the gamified enterprise system, on the right side of the screen, prominently positioned is the leaderboard listing top achievers in a number of categories. Some of the metrics used in calculating the position of employees on the leaderboard are through various mechanics, dynamics and collaborative or participatory endeavours.

The mechanics and rules which mediate the leaderboard system are such that all accumulated points are levelled at the end of the calendar year for every single employee in the company. This innovative game mechanic emerged from the design process, when the system developer and agency debated how to keep employees engaged throughout the leaderboard system. For example, when a number of highflyers were found to be consistently at the top of the leaderboard, this in turn proved challenging for other staff to gain recognition for their activities. Hitherto there was very little incentive for other staff to engage with the leaderboard system.

**Motivational Properties**

What transpires from our analysis, is that the motivational properties of the leaderboard is much higher for employees who engage with the gamified business system nearer the start of the calendar year. This is especially significant as this is when all active employees essentially commence their journey together through the digital infrastructure, as everyone's points begin variably at the same level. This provides an element of commonality and community, and reinforces the cultural ethos of 'we are in this together', which the agency tries to foster towards synergised collaboration through the business system. Everyone starts the game with the same abilities and opportunities at that stage, and the empirical evidence suggest that during this period, the motivational properties through the leaderboard system are very high. Individual employees are hoping to get recognised and gain kudos from their colleagues. However, when running the same analysis with employees who join the company at a later stage the findings reveal different engagement properties.

**Emerging Affordance**

The agency employs new staff at different times of the year, as it is sensible to recognise that businesses generally do not only recruit new staff at year start. Staffing is a constant organisational process due to factors such as expansion, new projects or natural attrition. When analysing specific interviews with employees who joined the agency mid-term, a common theme of disengagement with the leaderboard system emerges. For example, in one case an employee joined the company in September, at that stage the leaderboard highflyers were well established having earned points and rewards for their activities across the gamified business system during the previous eight months. In this case the employee's engagement with the leaderboard system was very low.

> "…there's no incentive for me to try and climb the leaderboard, have you seen where [name anonymised] is, look at how many [points] he has, how am I going to beat that?"

The challenge in earning enough points to get listed in any meaningful way on the system by employees joining mid-term, and therefore have their work activities and engagement with the gamified business system recognised, is perceived as being unsurmountable through the leaderboard. In this case a low level of engagement is afforded by properties of that game design element over time, which not only has the counterintuitive effect of disengaging certain members, but also impacts overall self-continuity of individual employees. Here, the leaderboard offers varying degrees of motivation and engagement depending on its temporal context rather than its situational setting.

This analysis into the leaderboard provides us with an insight into the emergent properties of game design elements from the design process, to the affordance between the employee and the gamified business system. These emergent properties are also found to originate from the design process, but also from the rules and engagement pull factor instantiated from the activities by colleagues within this business system.

### Temporal Dimension

Through our analysis the empirical data suggests temporal dimensions for game design elements in a gamified business system. Where a game element which provides the same function over time may be perceived differently and therefore affects the levels of motivation and ultimately the levels of employee engagement through the business system. This analysis offers implications in our understanding of the role and impact of game design elements for levels of employee engagement through a gamified business system.

## Conclusion

Although this research provides an insight into understanding game design element properties in a gamified business system, it is not without limitations. Using self-continuity, we analysed a specific game design element over time, however further research might engage in a wider analysis of congruent elements.

In this particular study, our findings suggest that game design elements exhibit emergent properties which can be situated in both the design and development process, but also by the instantiation from the activities of the users. These implication for practice provides a particularly salient insight into the emergent properties of game design element, where stakeholders tasked with developing specific elements of engagement as part of a wider business system, may not foresee the actual levels of engagement or motivational affordance. For research, this alludes to a continual and evolving process of examining properties of game design elements, and also how these can exhibit alternative emergent properties beyond their original design.

In summary, our contributions for knowledge in this research suggest that game design elements not only exhibit alternative emergent properties as a result of the interactions through the activities they support, but also how these mediate levels of motivation and engagement with a business system. This is prominent for practice, where organisations facing challenges in better engaging and motivating their workforce (Burke, 2014; Deterding, 2012; O'Boyle & Harter, 2014; Rauch, 2013), need to respond to system design challenges (Zhang, 2008). We therefore conclude by suggesting a dimension of emergence of game design elements, in the gamification of business systems to engage the workforce in the digital age.

# References

Adkins, A. (2015). Majority of U.S. Employees Not Engaged Despite Gains in 2014.   Retrieved 28 January, 2015, from http://bit.ly/1uUCjpX

Briscoe, G., & Marinos, A. (2009). *Digital ecosystems in the clouds: towards community cloud computing.* Paper presented at the Digital Ecosystems and Technologies, 2009. DEST'09. 3rd IEEE International Conference on. http://dx.doi.org/10.1109/DEST.2009.5276725

Burke, B. (2014). *Gamify: How gamification motivates people to do extraordinary things*: Bibliomotion, Inc. http://dx.doi.org/10.4324/9781315230344

Carayannis, E. G., & Rakhmatullin, R. (2014). The Quadruple/Quintuple Innovation Helixes and Smart Specialisation Strategies for Sustainable and Inclusive Growth in Europe and Beyond. *Journal of the Knowledge Economy, 5*(2), 212-239. http://dx.doi.org/10.1007/s13132-014-0185-8

Cheng, R. (2014). How Clash of Clans' clan wars got me addicted all over again.   Retrieved 26 April, 2014, from http://cnet.co/1HMpXZI

Ciborra, C. (2000). *From control to drift: the dynamics of corporate information infastructures*: Oxford University Press.

Ciborra, C., & Hanseth, O. (1998). From tool to Gestell: Agendas for managing the information infrastructure. *Information Technology & People, 11*(4), 305-327. http://dx.doi.org/10.1108/09593849810246129

Connor, H., Shaw, S., Shaw, S., & Fairhurst, D. (2008). Engaging a new generation of graduates. *Education+ Training, 50*(5), 366-378. http://dx.doi.org/10.1108/00400910810889057

Deci, E. L., & Vansteenkiste, M. (2004). Self-determination theory and basic need satisfaction: Understanding human development in positive psychology. *Ricerche di Psicologia*. http://dx.doi.org/10.1037/0003-066X.55.1.68

Deterding, S. (2012). Gamification: designing for motivation. *interactions, 19*(4), 14-17. http://dx.doi.org/10.1145/2212877.2212883

Deterding, S., Dixon, D., Khaled, R., & Nacke, L. (2011). *From game design elements to gamefulness: defining gamification.* Paper presented at the Proceedings of the 15th International Academic MindTrek Conference: Envisioning Future Media Environments. http://dx.doi.org/10.1145/2181037.2181040

El-Masri, M., Tarhini, A., Hassouna, M., & Elyas, T. (2015). *A design science approach to Gamify education: From games to platforms.* Paper presented at the Twenty-Third European Conference on Information Systems (ECIS).

Erturkoglu, Z., Zhang, J., & Mao, E. (2015). Pressing the Play Button: What Drives the Intention to Play Social Mobile Games? *International Journal of E-Business Research (IJEBR), 11*(3), 54-71. http://dx.doi.org/10.4018/ijebr.2015070104

Hamari, J., Huotari, K. and Tolvanen, J. (2015) Gamification from the Economics Perspective. In: Walz, S.P. and Deterding, S., Eds., The Gameful World: Approaches, Issues, Applications, MIT Press, Cambridge, 139-161.

Hamari, J., Koivisto, J., & Sarsa, H. (2014). *Does Gamification Work?—A Literature Review of Empirical Studies on Gamification.* Paper presented at the Proceedings of the 47th Hawaii International Conference on System Sciences. HICSS. http://dx.doi.org/10.1109/HICSS.2014.377

Herger, M. (2014). *Enterprise Gamification: Engaging People by Letting Them Have Fun*: CreateSpace Independent Publishing Platform.

Herzig, P., Ameling, M., & Schill, A. (2012). *A generic platform for enterprise gamification.* Paper presented at the Software Architecture (WICSA) and European Conference on Software Architecture (ECSA), 2012 Joint Working IEEE/IFIP. http://dx.doi.org/10.1109/WICSA-ECSA.212.33

Huotari, K., & Hamari, J. (2012). *Defining gamification: a service marketing perspective.* Paper presented at the Proceeding of the 16th International Academic MindTrek Conference. http://dx.doi.org/10.1145/2393132.2393137

Kapp, K. M. (2012). *The Gamification of Learning and Instruction: Game-based Methods and Strategies for Training and Education*: Wiley.

Khatib, F., DiMaio, F., Cooper, S., Kazmierczyk, M., Gilski, M., Krzywda, S., . . . Popović, Z. (2011). Crystal structure of a monomeric retroviral protease solved by protein folding game players. *Nature structural & molecular biology, 18*(10), 1175-1177. http://dx.doi.org/10.1038/nsmb.2119

King, D., Greaves, F., Exeter, C., & Darzi, A. (2013). 'Gamification': Influencing health behaviours with games. *Journal of the Royal Society of Medicine, 106*(3), 76-78. http://dx.doi.org/10.1177/0141076813480996

Landers, R. N., Bauer, K. N., Callan, R. C., & Armstrong, M. B. (2015). Psychological theory and the gamification of learning *Gamification in education and business* (pp. 165-186): Springer. http://dx.doi.org/10.1007/978-3-319-10208-5_9

Lee, Y.-T., Chen, K.-T., Cheng, Y.-M., & Lei, C.-L. (2011). *World of Warcraft avatar history dataset.* Paper presented at the Proceedings of the second annual ACM conference on Multimedia systems. http://dx.doi.org/10.1145/1943552.1943569

MacLeod, D., & Clarke, N. (2009). Engaging for success: enhancing performance through employee engagement: a report to government.

McGonigal, J. (2011). *Reality is broken: Why games make us better and how they can change the world*: Penguin.

Michael, D. R., & Chen, S. L. (2005). *Serious games: Games that educate, train, and inform*: Muska & Lipman/Premier-Trade.

Mollick, E. R., & Rothbard, N. (2013). Mandatory Fun: Gamification and the Impact of Games at Work. *The Wharton School Research Paper Series*. https://dx.doi.org/10.2139/ssrn.2277103

Monu, K., & Ralph, P. (2013). Beyond Gamification: Implications of Purposeful Games for the Information Systems Discipline. *arXiv preprint arXiv:1308.1042*.

Muntean, C. I. (2011). *Raising engagement in e-learning through gamification.* Paper presented at the Proc. 6th International Conference on Virtual Learning ICVL.

Nicholson, S. (2012). A user-centered theoretical framework for meaningful gamification. *Games+ Learning+ Society, 8*(1).

O'Boyle, E., & Harter, J. (2014). State of the global workplace: Employee engagement insights for business leaders worldwide: Gallup.

Paharia, R. (2013). *Loyalty 3.0: How to revolutionize customer and employee engagement with big data and gamification*: McGraw Hill Professional.

Penenberg, A. L. (2013). *Play at Work: How games inspire breakthrough thinking*: Hachette UK.

Petridis, P., Dunwell, I., De Freitas, S., & Panzoli, D. (2010). *An engine selection methodology for high fidelity serious games.* Paper presented at the Games and Virtual Worlds for Serious Applications (VS-GAMES). http://dx.doi.org/10.1109/VS-GAMES.2010.26

Rauch, M. (2013). Best practices for using enterprise gamification to engage employees and customers *Human-Computer Interaction. Applications and Services* (pp. 276-283): Springer. http://dx.doi.org/10.1007/978-3-642-39262-7_31

Reeves, B., & Read, J. L. (2013). *Total engagement: How games and virtual worlds are changing the way people work and businesses compete*: Harvard Business Press.

Richter, G., Raban, D. R., & Rafaeli, S. (2015). Studying Gamification: The Effect of Rewards and Incentives on Motivation *Gamification in education and business* (pp. 21-46): Springer. http://dx.doi.org/10.1007/978-3-319-10208-5_2

Ryan, R. M., & Deci, E. L. (2000). Self-determination theory and the facilitation of intrinsic motivation, social development, and well-being. *American Psychologist, 55*(1), 68. http://dx.doi.org/10.1037/0003-066X.55.1.68

Sani, F. (2010). *Self continuity: Individual and collective perspectives*: Psychology Press.

Singh, S. (2012). Gamification: A Strategic Tool for Organizational Effectiveness. *International Journal of Management, 1*(1), 108-113.

Skilton, M. (2015). *Building the Digital Enterprise*: Palgrave Macmillan. http://dx.doi.org/10.1057/9781137477729

Strader, T. J., Lin, F.-R., & Shaw, M. J. (1998). Information infrastructure for electronic virtual organization management. *Decision Support Systems, 23*(1), 75-94. http://dx.doi.org/10.1016/S0167-9236(98)00037-2

Supercell. (2014). How Often Do You Play Clash Of Clans? (On Average A Day).   Retrieved 7 December, 2014, from http://bit.ly/1ML7s97

Thom, J., Millen, D., & DiMicco, J. (2012). *Removing gamification from an enterprise SNS.* Paper presented at the Proceedings of the ACM 2012 conference on Computer Supported Cooperative Work. http://dx.doi.org/10.1145/2145204.2145362

Wood, R. T., Griffiths, M. D., & Parke, A. (2007). Experiences of time loss among videogame players: An empirical study. *CyberPsychology & Behavior, 10*(1), 38-44. http://dx.doi.org/10.1089/cpb.2006.9994

Zhang, P. (2008). Technical opinion Motivational affordances: reasons for ICT design and use. *Communications of the ACM, 51*(11), 145-147. http://dx.doi.org/10.1145/1400214.1400244

Zichermann, G., & Cunningham, C. (2011). *Gamification by Design: Implementing Game Mechanics in Web and Mobile Apps*: O'Reilly Media.